\documentclass{an}
\usepackage{graphicx}
\usepackage{times}
\usepackage{fancyhdr}
\begin{document}

\title{EM Cygni: a study of its eclipse timings}

\author{Sz. Csizmadia$^1$, Zs. Nagy$^2$, T. Borkovits$^3$, T. Heged\"us$^3$, 
        I. B. B\'{\i}r\'o$^3$, Z. T. Kiss$^3$}
\institute{Konkoly Observatory of the Hungarian Academy of Sciences, 
           H-1525 Budapest, P. O. Box. 67., Hungary, e-mail: csizmadia@konkoly.hu
	   \and
	   Roland E\"otv\"os University,
           H-1121 Budapest, P\'azm\'any P. stny. 1/A, Hungary
	   \and
	   Baja Astronomical Observatory of B\'acs-Kiskun County, 
	   H-6500 Baja, Szegedi \'ut, P. O. Box 766,  Hungary
	   }
\date{Received; accepted; published online}

\abstract{EM Cygni is a Z Cam-subtype eclipsing dwarf nova. Its
          orbital period variations were reported in the past but
          the results were in conflict to each other while other
          studies allowed the possibility of no period variation.
          In this study we report accurate new times of minima
          of this eclipsing binary and update its $O-C$ diagram.
          We also estimate the mass transfer rate in EM 
	  Cygni system and conclude that the mass transfer is far 
	  from the critical value. The mass transfer rate determined 
	  from the eclipse timings is in agreement with the 
	  spectroscopically determined value.}


\authorrunning{Csizmadia et al.}

\maketitle

\section{Introduction}

Dwarf novae are binary stars consisting of a white dwarf and a red dwarf
in which some mass is transferred from the red dwarf component onto the
white dwarf. Depending mainly on the mass transfer rate and the magnetic
field of the white dwarf they produce outbursts caused by instabilities
in the accretion disk. In the case of weak magnetic field of the white
dwarf the main factor which governs the frequency and amplitude of the
outbursts is the mass transfer rate. There is a critical value of the
mass transfer rate and it was investigated by Shafter et al.  (1986).
According to their results, if the mass transfer rate is higher than
this critical value there are no dwarf nova eruptions (we have a
nova-like variable) and if the mass transfer rate is below it we have a
dwarf nova. If the mass transfer is nearly equal to the critical value
we have a Z Cam-type variable (see Shafter et al. 1986, 2005). (For
details on dwarf novae and the processes in them see Warner 1995).

In the General Catalogue of Variable Stars (Samus et al. 2004) sixteen 
eclipsing dwarf novae were listed. These objects are important because the 
period can easily and precisely be measured due to the eclipsing nature 
and this allows to demonstrate so far small astrophysical effects as mass 
transfer rate. The careful analysis of the eclipse timings can reveal 
small orbital period changes which cannot be followed by radial velocity 
studies. For details on the eclipse geometry in dwarf novae see Smak 
(1994).

The subject of this study is EM Cygni which is the only known eclipsing Z
Cam-subtype dwarf nova{\footnote{The eclipsing dwarf nova V729 Sgr
may also belong to the Z Cam-subtype, but the identification of its
subtype is uncertain (Cielinski et al. 2000).}}. Therefore the precise
determination of its mass transfer rate is extremely important. Its
eclipses were discovered by Mumford and Krzeminski (1969) and this allows
to determine its mass transfer rate. Pringle (1975) and Mumford (1980)
stated that the eclipse period of EM Cyg is decreasing. In contradiction
with these results, Beuermann \& Pakull (1984) concluded that there is no
evidence for period change in EM Cyg. Herczeg (1987) also investigated the
period of EM Cygni and found weak evidence for period change but he
concluded that more observations are needed. It is worthy to mention that
the period decrease rate determined by Pringle (1975) was in the order of
$10^{-10}$ days/cycle while Mumford's (1980) value is in the order of
$10^{-11}$ days/cycle. These rather contradictory results may originate 
from the fact that the precision of minimum times obtained before 1984
(Beuermann and Pakull 1984 estimated their accuracy to be 0.0008 days)
could be compared to the expected width of the $O-C$ in such a small
period variation.  If the time-coverage of the minimum observations would
be longer, this effect can be avoided because the accumulated period
variations are larger in the $O-C$ than its intrinsic scatter.

Since 1984 more than 25000 revolutions have occurred and the corresponding 
$O-C$ value -- assuming $10^{-12}$ days decreasing in every cycle -- reaches 
$-0.0036$ days. Twenty-one years after the latest published photoelectric 
minimum in any refereed journal we decided to observe EM Cygni because the 
accumulation of the small continuous period changes can be sufficient
for deciding whether the EM Cygni's period is variable or not. This is very 
important from the point of view of understanding the mass transfer 
mechanism in cataclysmic variables.

Our aim was twofold. First, we wanted to solve the question of 
period change of EM Cyg with new observations. Secondly, we wanted to 
calculate the mass transfer rate from the observed period change if such a 
change is present.

\section{Observations}

\object{EM~Cyg} was observed at the Konkoly Observatory on four nights 
(in Cousins I-band) and at the Baja Astronomical Observatory on two nights 
(in V-band) to obtain precise light curves. The observations were carried 
out with the 1m RCC telescope of the Konkoly Observatory located at the 
Piszk\'estet\H o Mountain Station at 964 meters above the sea level and 
with the 50cm robotic telescope of the Baja Astronomical Observatory.

The detector of the 1m RCC telescope was an 1340$\times$1300 pixels
electronically cooled  ($T_{CCD} = -40^\circ$C) CCD camera manufactured by the
Roper Scientific  Ins. The exposure time was between 5-15 seconds depending on 
sky-conditions. The readout-time was only two seconds therefore the duty cycle
was very favourable. All frames were bias-subtracted. Dark  correction was not
applied because a negligible dark current was measured  for this exposure time.
Flat-field correction was also done using dome  flats. The detector of the
50cm robotic telescope was an Apogee ALTAU16 4kx4k CCD camera with a field of
view of $42\times42$ arcminutes. Bias, dark and flat-field corrections were
applied. To determine the raw magnitudes of stars in the frames, aperture 
photometry was performed using the IRAF/DAOPHOT package (Stetson 1990). We 
selected a comparison star and a check star in the frames and they showed  a
sufficient stability in each colour (standard deviations of their  magnitude
differences is 0.003 magnitudes (1m RCC) and 0.007 magnitudes (50cm robotic telescope),
respectively). Then differential photometry  was done
defining the variable's brightness as $\Delta m = m_{var} -  m_{comp}$. No
standard transformation was applied.

\section{Analysis of the $O-C$ diagram}

\subsection{Source of data}

Times of eclipses were collected from the following sources. 29 minima 
times were taken from Mumford \& Krzeminski (1969); two from Mumford 
(1974); one from Mumford (1975); three from Mumford (1980). One time of 
minimum was observed by Beuermann \& Pakull (1984). These are all the 
published minima in any refereed journal and none has been published since 
them.

Beuermann \& Pakull (1984) determined two times of minima from the Figure 1 of Jameson et
al. (1981). Of course, this method does not yield an  accurate minimum time, but we followed
them and concluded that around JD  2444105 -- when the observations of Jameson et al. (1981)
were carried out  -- the $O-C$ value was about $-0.0065$ days (against the ephemeris of
Mumford \& Krzeminski 1969). (At this time the cycle number  $E$ was close to 21390.) We
assign a one-tenth weight to this value.

We also found two amateur minimum observations on the homepage of 'The 
Astronomer' (www.theastronomer.org). One of them cannot be used for a 
minimum time estimation but the other one -- reproduced in Figure~1 -- 
was inspected and a fitting resulted a usable minimum time. We denoted 
the reference of this minimum time as 'Anonymous' in Table 1.

New and reliable minima times were determined from our light curves with the
Kwee-van Woerden (1956) method which means that the descending and the ascending
branches of the minimum are reflected to a  certain time value and it searches
for the time which yields the minimum  difference between the original and the
reflected curves. These moments of minima are presented in Table 1.  Errors of
the times of minima can be estimated as follows: 0.0008 days before 1980,
0.0001-0.0004 after that. These errors are smaller than the observed scatter of
$0.002$ days seen in Figure~4 (which shows the $O-C$ diagram). The possible
explanation of this can be due to the intrinsic variations of the location of
the hot spot in the system. Note that the eclipses in EM~Cyg are not total but
grazing ones. Some of our minimum observations are presented in
Figures~2-3.

\begin{table}[h]
\caption{Times of minima of EM Cygni. References: (1): Mumford \& Krzeminski 
(1969), (2): Mumford (1974), (3): Mumford (1975), (4): Mumford (1980), (5): 
Beuermann \& Pakull (1984), (6): Anonymous (see text), (7): this paper.}
\label{minima}
\begin{tabular}{lrlr}\hline
HJED        & Reference & HJED & Reference \\
\hline
37882.8603 & 1 & 39230.9335 &1\\
37883.7321 & 1 & 39293.7704 &1\\
37906.7130 & 1 & 39767.6624 &1\\
37911.6603 & 1 & 39769.6971 &1\\
37936.6778 & 1 & 40006.7886 &1\\
37966.6413 & 1 & 40007.9534 &1\\
37968.6778 & 1 & 40008.8263 &1\\
37996.6048 & 1 & 41980.6054 &2\\
38174.9335 & 1 & 41982.6435 &2\\
38345.6984 & 1 & 42515.8792 &3\\
38348.6058 & 1 & 43776.6771 &4\\
38496.9701 & 1 & 43778.7146 &4\\
38561.5523 & 1 & 43780.7508 &4\\
38562.4242 & 1 & 45257.4091 &5\\
38624.3885 & 1 & 50692.4613 &6\\
38674.7156 & 1 & 53650.4266 &7\\
38675.5883 & 1 & 53709.1873 &7\\
38676.7513 & 1 & 53984.3938 &7\\
38878.9343 & 1 & 53989.3343 &7\\
38883.8795 & 1 & 53990.5006 &7\\
39052.6043 & 1 & 53991.3676 &7\\
39054.6428 & 1 & 53993.4088 &7\\
\hline
\end{tabular}
\end{table}

\begin{figure}
	  \resizebox{\hsize}{!}
{\includegraphics[]{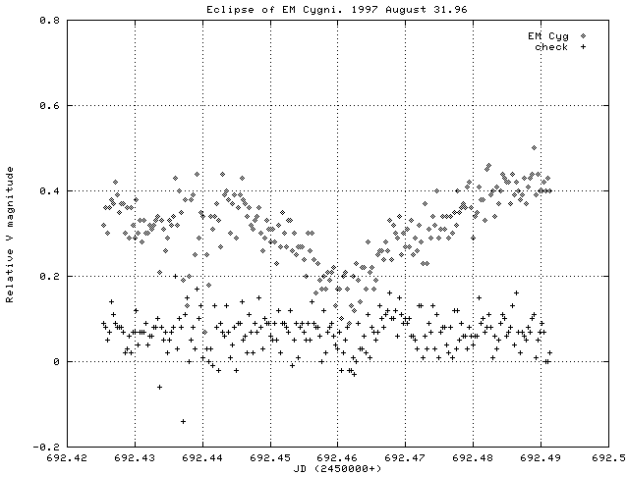}}
\caption{The minimum observations of EM Cyg reproduced from the homepage of {\it The Astronomer}.}
\label{fig1}
\end{figure}

\begin{figure}
\resizebox{\hsize}{!}
{\rotatebox{-90}{
{\includegraphics[]{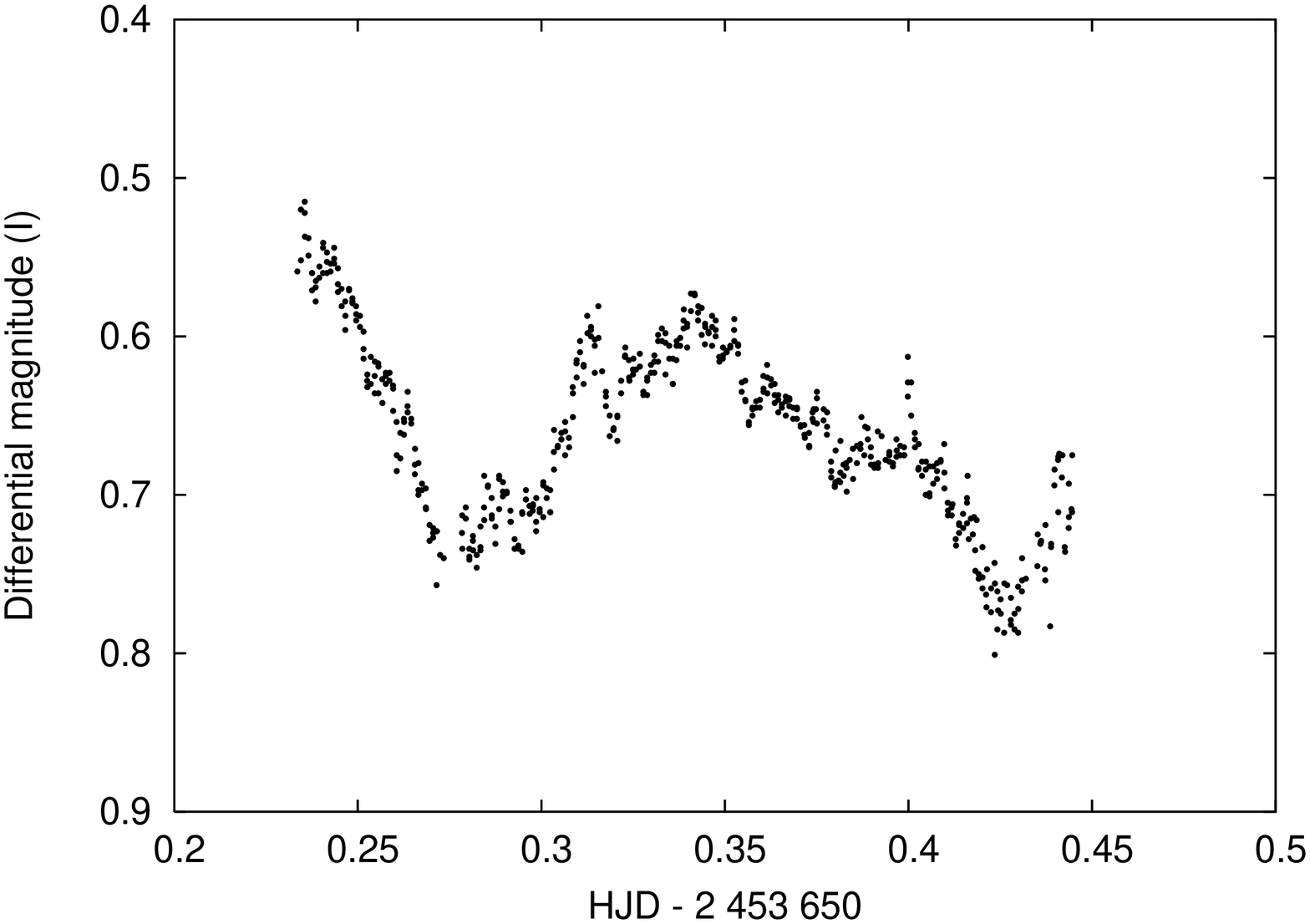}}}}
\caption{The I-band light curve of EM Cyg obtained by us \newline
 at JD 2 453 650.}
\label{fig2}
\end{figure}

\begin{figure}
\resizebox{\hsize}{!}
{\rotatebox{-90}{
{\includegraphics[]{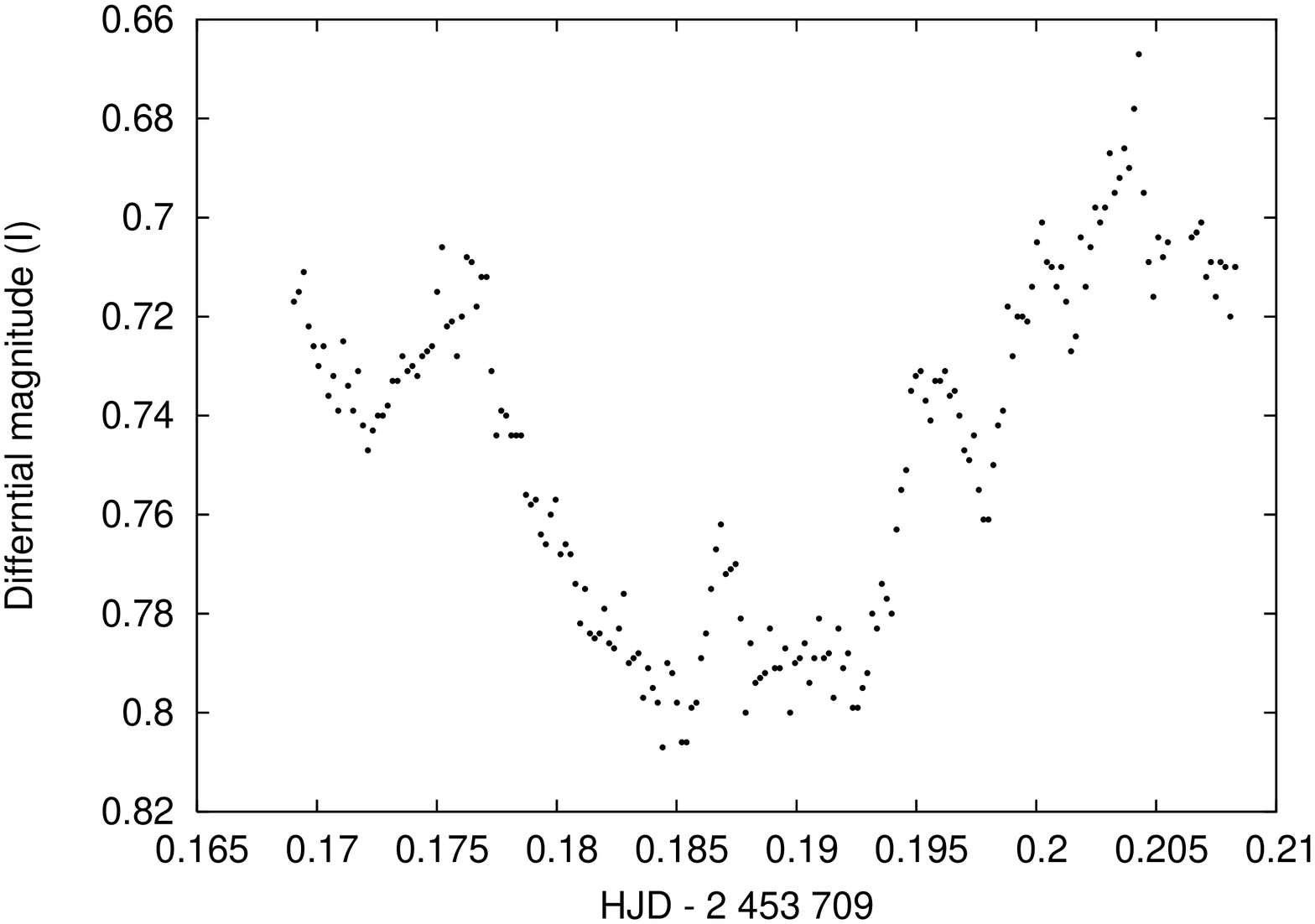}}}}
\caption{The I-band light curve of EM Cyg obtained by us \newline
at JD 2 453 709.}
\label{fig2}
\end{figure}

\begin{figure}
\resizebox{\hsize}{!}
{\rotatebox{-90}{
{\includegraphics[]{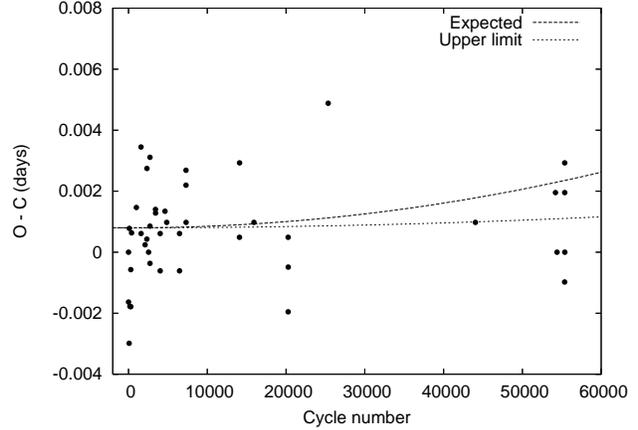}}}}
\caption{$O-C$ diagram of EM Cyg. The 'expected' is the theoretically 
expected $O-C$ curve calculated from the theory (Eq. (8) of Shafter et al. 1986) 
assuming conservative case. The 'upper limit' curve shows the maximum possible 
period variation (see text).}
\label{fig3}
\end{figure}

\subsection{The $O-C$ diagram}

The Julian Heliocentric Ephemeris Date (HJED) is free from the varying 
rotation of Earth. Since we need as high precision as we can reach we 
transformed all HJD values of the observed minima into HJED ones. Then  
linear and quadratic fits yield the following ephemeris:
\begin{equation}
T_{\rm ecl}(\rm HJED) = 2437882.8606(3) + 0.29090912(1) \times E 
\end{equation}
and
\begin{eqnarray}
\lefteqn{T_{\rm ecl}(\rm HJED) = 2437882.8606(3) + 0.29090911(4) \times E }\nonumber \\
& - 2(2) \cdot 10^{-14} \times E^2
\label{eph2}
\end{eqnarray}
The numbers in the paranthesis show the errors in the last digits.

We also calculated the corresponding $\chi^2$ values for both fits and  found
that in case of the linear fit $\chi^2$ = 1.99 while in case of  quadratic fit
it is $\chi^2$= 1.98. So the introduction of a quadratic  term does not improve
significantly the fit. However, a simple least-squares regression could be 
very sensitive to the distribution of data points and of errors. Therefore we
determined the value of maximum possible quadratic term in another way. We
calculated the $O-C$ values within a parameter-space  (in the form $T_{ecl} =
T_0 + A E + B E^2$ where $E$ is the cycle number, and $T_0 = 2~437~882.8590 ...
2~437~882.8622$, step size: $0.0001$ days, $A = 0.29090910 ... 0.29090915$, step
size: $2.4\cdot10^{-8}$ days, $B = -1.5 \cdot 10^{-12} ... 1.5 \cdot 10^{-12}$,
step size: $10^{-15}$ days) and we found that the corresponding $\chi^2$ value
of the $O-C$ values are minimal at $T_0 = 2~437~882.8598$, $A = 0.29090914$, $B =
-9.8 \cdot 10^{-14}$. This corresponds to $\dot P/P < 2.32 \cdot 10^{-12}$
1/days. A Monte Carlo simulation was also used to estimate the upper limit
because in this way we could study how the errors may yield an apparent
quadratic term. At the epochs of the observed $O-C$ curve, 1000 $O-C$ values 
were simulated for each epoch with random errors. The mean of these errors 
was as mentioned in the previous section and they were assumed to be  normally
distributed. Then we had 1000 $O-C$ curves and all of them were fitted by a
parabola. The average of the quadratic terms yielded $<B> = -8.0 \cdot 10^{-14}$
days. This is similar to the result of the previous attempt. Therefore 
the value of $2.32 \cdot 10^{-12}$ 1/days was accepted as the absolute value of
the upper limit for the period decrease.

For comparison, we plotted the $O-C$ diagram of EM Cygni calculated by the
linear ephemeris together with the theoretically expected one in Figure~4. 
The maximum possible period variation is also shown in Figure~4
which was calculated from the above determined upper limit of $\dot P$.

Comparing our ephemeris to the previously determined ones (see Section~1)
one can see that our result is close to the one of Beuermann and Pakull (1984):
we found no evidence of period variation. We cannot confirm the results of
Pringle (1975) and Mumford (1980) who found definite period decrease.

\section{Discussion}

\subsection{Conservative case}

Since we could only give an upper limit for the period variation only, we could guess
the maximum possible mass transfer rate in the system. Assuming a conservative case 
(i. e. the angular momentum and total mass remains constant in the system) the mass 
transfer rate can be related to the period change via
\begin{equation}
\frac{1}{P}\frac{dP}{dt} = -3\left(\frac{1}{q} - q\right) \frac{1}{M_1 + M_2} \frac{d{M_2}}{dt}
\end{equation}
(e. g. Thomas 1977) where $t$ refers to time, $q$ is the mass ratio and
{\bf $M_2$} is the mass of the donor star. Accepting $q=0.88$ and {\bf 
$M_1 = 1.12M_\odot$}, $M_2 = 0.99M_\odot$ (North et al. 2000) and 
$P^{-1} dP/dt = 2.32\cdot10^{-12}$ we have $dM_2/dt = 2 \cdot
10^{-9}M_\odot$/yr. This is an upper limit for the mass transfer rate.

From the study of the far-ultraviolet spectrum of EM Cygni, Winter \& 
Sion (2003) found that the mass transfer rate in EM Cyg is between 
$3.2\cdot 10^{-11}$ and $1.1\cdot 10^{-10}$ solar mass per year. Our 
upper limit is higher than their upper value and is not in contradiction 
with their results. Therefore the agreement between the spectrocopically 
observed mass transfer rate and the one derived from eclipse timings is 
good.

We compared the result to the theoretical expectations. Shafter et al. (1986)
investigated the value of the critical mass transfer rate for  different values
of orbital periods and white dwarf masses. Accepting  $M_1=1.12 M_\odot$ for the
white dwarf in EM Cyg (North et al. 2000) and  taking into account that EM Cyg
has an orbital period of 6.96 hours, Eq. (8) of Shafter et al. (1986) yields
$dM_{crit}/dt =  1.2\cdot10^{-8}M_\odot$/yr. If the mass transfer is
conservative in EM  Cyg, this system is far from this limit. In summary, the
mass transfer  rate is lower by an order than the theoretically expected
value.

\subsection{Non-conservative case}

In EM Cyg the mass is transferred from the less massive component to the 
more massive one yielding a period increase. In non-conservative mass 
transfer there are several mechanisms which decrease the period. 
Therefore, in total, the result can be a nearly constant as well as a 
decreasing period! In this case the period variation depends on the mass 
transfer rate as well as the angular momentum ($J$) variations: 
\begin{equation}
\frac{\dot P}{P} = 3 \frac{\dot J}{J} - \frac{2+3q}{1+q}\frac{\dot M_1}{M_1} - \frac{3+2q}{1+q}\frac{\dot M_2}{M_2} 
\end{equation} 
(see Warner 1995). Assuming that $\dot M_2 = -\dot M_1$ we have 
\begin{equation} 
\frac{\dot P}{P} = 3 \frac{\dot J}{J} + \frac{3(1-q^2)}{q(1+q)}\frac{\dot M_1}{M_1} \end{equation} 
We observed nearly zero period variation so we can determine what 
angular momentum loss is required to neutralize the period increase 
caused by the mass transfer. For this purpose we set $\dot P = 0$ 
and $\dot M_1$ was chosen as the critical value. With $q=0.88$, $M_1 = 
1.12 M_\odot$ and $\dot M_1 = 1.2\cdot 10^{-8}M_\odot$/yr we get that 
$\frac{\dot J}{J}$ are to be $-4.63\cdot 10^{-17} {\rm s}^{-1}$ in order 
to reduce the period variation to zero.

The system can loss mass via stellar wind. Assuming that the secondary 
loses $10^{-14}$ solar mass per year via stellar wind (which 
is a similar value to the one of the Sun) and all this mass isotropically 
leaves the system, this leads $\dot P / P = -2.4\cdot 10^{-18}{\rm 
day}^{-1}$ (we used the formula given in Batten 1973, p.95). This 
is too low to explain the missing period variation.

The rotational period of a rotating star decreasing, if the star loses mass by
stellar wind (Iben, Fujimoto \& MacDonald 1992). Due to the
spin-orbit coupling this reduces the total angular momentum of the system. From
Warner's (1995) Eq. (9.13b) we know that
\begin{equation}
\dot J_{\rm rot} = -1.2 \cdot 10^{34} \left(\frac{k_{2}^{2}}{0.1}\right)^2 P_{orb}^{31/12}(h) \mathrm{dyncm}
\end{equation}
In this relation $k_2$ is the gyroradius of the secondary. Assuming $k_2 = 0.1$
we have $\dot J / J = -2.25 \cdot 10^{-16}{\rm s}^{-1}$ due to this rotational
angular momentum loss. This seems to be too high to explain the constancy of the
period.

The gravitational radiation is negligible in such systems. The magnetic braking
can be a more efficient mechanism. This results in angular momentum loss on the
scale of
\begin{equation}
\dot J = -2.52 \cdot 10^{34} P_{orb}^{1.64} (h) \mathrm{dyncm}
\end{equation}
(McDermott \& Tamm 1989, Warner 1995). In EM Cygni this yields $\dot J/J = -7.57
\cdot 10^{-17}{\rm s}^{-1}$. This is higher than the required value by a factor
of 2, but regarding the rather large uncertainties of magnetic braking
theories and the fact that we substitute $\dot P / P = 0$ instead of its correct
value (which is too low to determine it exactly) one can conclude that this
itself could explain the constancy of the period.

To hold up the mass transfer some angular momentum loss is required 
(Warner 1995). Using Warner's (1995) Eq. (9.16) we found that this means 
$\dot J / J = -1.2\cdot10^{-16}{\rm s}^{-1}$ which is higher by a 
factor of 2.6 than required.

One can assume that more than one of these mechanisms are working in this
system. But the sum of them is higher than the required value. If the mass
transfer rate is equal to the prediction of Shafter et al. (1986), then the
observed constancy of the period can be explained by the magnetic braking
itself, and the uncertainty of the estimations allows that we can regard the
effects of mass transfer and the magnetic braking to be equal. Also,
magnetic braking can be the process which drives the mass-transfer (because its
magnitude is in the required range). But in this case
Winter \& Sion (2003) have been measured too low mass transfer rate.

The mass transfer rate given by Winter \& Sion would mean an 
angular momentum loss rate of $\dot J / J = -2.7 \cdot 10^{-19} {\rm 
s}^{-1}$ if the period variation is zero. This is close to the range of 
the angular momentum loss caused by the assumed value of the secondary's 
stellar wind rate (note that this real or accidental agreement does not 
mean that this should be due to the angular momentum loss caused by 
stellar wind) but it does not fit the rate of other kind of possible 
angular momentum loss. With their mass transfer rate the magnetic 
braking and/or the rotational braking would dominate the right side hand 
of Eqs. (4-5) and we would observe a large period decrease. One can estimate the
period variation rate from their $\dot M$ value. Substituting 
their mass transfer rate into Eq. (5) and assuming that there is an angular 
momentum loss from magnetic braking (Eq. 7) we would have $\dot P / 
P = -2.0\cdot 10^{-11} {\rm days}^{-1}$. From the available 
minima observations this figure should be excluded.

\section{Conclusion}

We observed six new times of minima of EM Cygni. We updated its $O-C$ 
diagram and found that the period is constant. We found that the 
upper limit for the period decrease is $|\frac{\dot P}{P}| < 2.3 \cdot 
10^{-12}$ 1/days.

If the mass transfer is conservative then its rate is far from the 
theoretically predicted value (Sect. 4.1) but it is in agreement of the 
spectroscopically determined mass transfer rate of Winter and Sion 
(2003). (Note that their method was indirect and its reliability 
depends on e.g. the accuracy of distance therefore this agreement 
between the two independent determinations give a hint only that the 
mass transfer is not conservative.) In conservative case the 
spectroscopically observed mass transfer rate and the one given by 
present eclipse timing analysis is lower than the theoretically 
predicted value.

If we assume that the mass transfer occurs in a non-conservative mode 
and the rate of mass transfer equals to the theoretical predictions of 
Shafter et al. (1986), we found that the period increase caused by the 
mass transfer from the secondary to the primary is reduced by the 
angular momentum loss due to magnetic braking to zero (Sect. 4.2). This 
explains why we did not observe period variation. It is worthy to 
mention that in this case the spectroscopic measurement of the mass 
transfer rate by Winter \& Sion (2003) is in contradiction.

Therefore their spectroscopic measurement would exclude the 
scenario we investigated in Sect. 4.2. If their measurement is right 
then the mass transfer cannot occur in non-conservative case   
and this would mean that the accretion disk theory model of dwarf 
novae may be challenged by the currently available observations.

Since EM Cyg is the only known eclipsing Z Cam-subtype dwarf nova its further
minima observations are important to refine our results and determine more
precisely the difference between the observed and expected mass transfer rates.

\acknowledgements
We thank the referee for his/her suggestions which improved significantly the 
paper.


\end{document}